\documentclass[10pt,twocolumn,twoside] {IEEEtran}

\usepackage{cite}

\ifCLASSINFOpdf

\else
   \usepackage[dvips]{graphicx}
   \DeclareGraphicsExtensions{.eps}
\fi
\usepackage[cmex10]{amsmath}
\usepackage{array}
\usepackage{amsthm}
\theoremstyle{remark}

\newtheorem{remark}{Remark}
\newtheorem{proposition}{Proposition}
\usepackage{booktabs}
\usepackage{bm}
\usepackage{tabularx}

\usepackage{color} 
\usepackage{booktabs}
\usepackage{bm}
\usepackage{tabularx}
\usepackage{amssymb} 
\usepackage{algpseudocode}  

\begin{document}
\title{ Joint DOA and Frequency Estimation\\ with Sub-Nyquist Sampling}
\author{Liang~Liu
        and~Ping~Wei%
\thanks{The authors are with the Center for Cyber Security, School of Electronic Engineering, University of
Electronic Science and Technology of China, Chengdu, 611731 China (e-mail:liu\_yinliang@outlook.com; pwei@uestc.edu.cn).}}

\markboth{Journal of \LaTeX\ Class Files,~Vol.~11, No.~4, December~2012}%
{Shell \MakeLowercase{\textit{et al.}}: Bare Demo of IEEEtran.cls for Journals}

\maketitle

\begin{abstract}
In this paper, to jointly estimate the frequency and the direction-of-arrival(DOA) of the narrowband far-field signals, a novel array receiver architecture is presented by the concept of the sub-Nyquist sampling techniques. In particular, our contribution is threefold.
i) First, we propose a time-space union signal reception model for receiving array signals, where the sub-Nyquist sampling techniques and arbitrary array geometries are employed to decrease the time-domain sampling rate and improve the DOA estimation accuracy.  A better joint estimation is obtained in the higher time-space union space.  ii)  Second, two joint estimation algorithms are proposed for the receiving model. One is based on a trilinear decomposition from the third-order tensor theory and the other is based on subspace decomposition. iii) Third, we derive the corresponding Cram\'{e}r\text{-}Rao Bound (CRB) for frequency and DOA estimates.
In the case of the branch number of our architecture is equal to the reduction factor of the sampling rate, it is observed that the CRB is robust in terms of the number of signals, while the CRB based on the Nyquist sampling scheme will increase with respect to the number of signals. In addition, the new steer vectors of the union time-space model are completely uncorrelated under the limited number of sensors, which improves the estimation performance. Furthermore, the simulation results demonstrate that our estimates via the receiver architecture associated with the proposed algorithms closely match the CRB according to the noise levels, the branch number and the source number as well.
\end{abstract}


\IEEEpeerreviewmaketitle

\section{Introduction}
\IEEEPARstart{J}{oint} estimation of carrier frequency and direction of arrival (DOA) for multiple signals {is desired} in many practical applications. For example, Cognitive Radio (CR) technique might be a good way to cope with the problem of the spectral congestion~\cite{Haykin2005, Yucek2009, Mishali2011a, Sun2013, Cohen2014}. One of the most important functions of CRs is to detect locally idle spectrum and then make the spectrum access from the concept of spectrum sensing. Generally, {there are three dimensions of spectrum space, i.e., time, frequency and space}. With the development of array processing techniques~\cite{Krim1996, Schmidt1986, Roy1986}, the spatial spectrum or DOA of a signal can be thought as a new approach to improve the performance of CRs. Therefore, { more effort have been spent on } how to jointly estimate carrier frequencies and their DOAs of multiple signals~\cite{Lemma1998, Lemma2003}. Unfortunately, both of them exist at least two shortcomings. One is the pair matching problem for the carrier frequencies associated with the DOAs. The other is that time domain sampling rate is equal to or larger than the Nyquist sampling rate which is considered as a bottleneck for wideband signal processing by CRs. For instance, it leads to prohibitive Nyquist sampling rate and massive sampling data to be processed if the spectrum needed to be monitored from 300 MHz to several GHz~\cite{Haykin2005, Yucek2009, Mishali2011a, Sun2013, Cohen2014}.

To deal with the problem of high sampling rate, recently, the sub-Nyquist sampling technique has been proposed to reconstruct a multiband signal from the data obtained under the Nyquist sampling rate~\cite{Mishali2011,Mishali2010,Eldar2009,Mishali2009}. Inspired by the idea, some methods were presented for the joint estimation of DOA and carrier frequency based on sub-Nyquist sampling rates. The authors of~\cite{Ariananda2013} suggested a new structure, where each output of a linear array is carried out through the multi-coset sampling. {In~\cite{Ariananda2013}  the minimum redundancy array (MRA) is employed to estimate the DOA of more uncorrelated sources than sensors.} In this way, the wide-sense stationary signal can be compressed in both the time domain and the spatial domain. {The frequency and DOA estimation accuracy are limited by the reciprocal of block length and array aperture, respectively. And it need a two-dimensional (2D) peak searching to get the frequency and DOA estimation from the 2D power spectrum.}
To simplify the hardware complexity, an additional identical delayed channel for each antenna is suggested in~\cite{Kumar2014}. Herein, the problem of pairing ambiguity will {arise} using an underlying uniform linear array (ULA). And then~\cite{ Kumar2015} proposed a structure, which has the same hardware complexity as that of~\cite{Kumar2014}. In ~\cite{Kumar2016}, the authors proposed the so-called space-time array to jointly estimate frequency and DOA when the number of sources is more than the number of sensors. {However, those methods in~\cite{Kumar2015,Kumar2016}  are limited to ULA because they make use of the  rotation invariance property of ULA.}
More recently, two joint DOA and carrier frequency recovery approaches based on the L-shaped ULAs are presented in~\cite{Stein2015}. However, all of these papers did not give a unified signal reception model for the array receivers.

{Dealing with the problem of joint frequency and DOA  estimation, it is widespread that the spatial samplings are less because of limited sensors number, and the temporal samplings are enough. A kind of very natural viewpoint is jointly considering in both time domain and space domain. If we unite time and space domain through elaborately modeling,  we will have more chances to classify targets. Because the differences between vectors from not only spatial space but also temporal space will be reserved in the new  union space, besides, the differences between vectors in the new union space will be enlarged even if the differences between vectors from any one space are small,  we employ Kronecker or Khatri-Rao product to jointly estimate the frequency and DOA from the time-space union space in this paper. Because the new vectors in the union space have bigger differences and less correlation, we can better classify the targets based on the differences from not only  frequency domain but also spatial domain.}

{In this paper, a new array receiver architecture is proposed. Associated with two sub-Nyquist sampling based methods, we simultaneously estimate the frequencies and DOAs of multiple narrowband far-field signals impinging on a array, where signals' carrier frequencies spread around the whole wide spectrum. {It is noteworthy that the array is not limited to ULA. The other arrays can be applied to gain their advantage, such as MAR can achieve a higher estimation accuracy than ULA with same sensor number.}  Since the reception model of our receiver makes use of the result on Kronecker product, the joint DOA and frequency estimation will benefit from it.} In addition, the Cram\'{e}r\text{-}Rao Bound (CRB) for spatial phase estimation is also derived based on this model. It is proven that the CRB is not {affected} by the signal number when the branch number of our architecture is equal to the sampling rate reduction factor, while the CRB using Nyquist sampling will increase according to the signal number. In other words, our model's CRB is lower than the CRB which employs Nyquist sampling. Finally, the simulations confirm the above conclusion on CRB and the superior performance of the proposed methods from three aspects: noise level, the number of branch, and the number of source as well.

This paper is organized as follows: in Section II, we describe the basic array signal model and point out the objective of this paper. In Section III, the proposed receiver architecture is presented, and a new  signal reception model is derived. In Section IV, two joint DOA and frequency estimation methods for the receiver architecture are proposed. In Section V, we deduce the corresponding CRB and demonstrate the result on CRB. Section VI carries out the simulation experiment and finally the conclusions of this paper are given in Section VII.

The following notations are used in the paper.  ${\left(  \cdot  \right)^{\rm T}}$, ${\left(  \cdot  \right)^{\rm H}}$, and ${\left(  \cdot  \right)^{^\dag }}$  denote the transpose, Hermitian transpose, and {Moore-Penrose pseudo-inverse}, respectively. $E\left(  \cdot  \right)$ stands for the expectation operator.  ${x_j}$ is the  $j$th entry of a vector ${\bf{x}}$.  ${{\bf{A}}_i}$, ${{\bf{A}}^j}$, and ${A_{ij}}$ are the $i$th row, the $j$th column, and $(i,j)$th entry of a matrix ${\bf{A}}$, respectively. $ \otimes$, $\odot$, and $\ast$  denote the Kronecker product, Hadamard product, and Khatri-Rao product, respectively.  ${{\bf{I}}_M}$ stands for an $M \times M$ identity matrix.

\section{Signal model and objective}
In this section, we will give the array signal model and fundamental assumptions as well as the objective of this paper.

\subsection{Array signal model}
Consider $K$ narrowband far-field signals impinging on an array composed of $M$ $(M>K)$ sensors. It is assumed that the signals' center frequencies are separate widely. Thus, the narrowband far-field signals can be modeled as multiband signals in \cite{Mishali2009}.  The array output can be written as\cite{Krim1996}
\begin{align}\label{eqnArray}
{\bf{x}}\left( t \right) = {\bf{As}}\left( t \right) + {\bf{n}}\left( t \right),
\end{align}
where ${\bf{x}}\left( t \right) = {\left[ {{x_1}\left( t \right), \cdots ,{x_M}\left( t \right)} \right]^{\rm{T}}}$ is the measurement vector, ${\bf{s}}\left( t \right) = {\left[ {{s_1}\left( t \right), \cdots ,{s_K}\left( t \right)} \right]^{\rm{T}}}$ is the vector of all signal values, where the signals are uncorrelated, ${\bf{n}}\left( t \right)= {\left[ {{n_1}\left( t \right), \cdots ,{n_M}\left( t \right)} \right]^{\rm{T}}} $ { is the zero-mean complex spatially and temporarily white Gaussian noise vector, whose variance is  ${\sigma}^2$}.
{As the most widely used array, who also derive many non-uniform linear array, the ULA is taken into consideration. The array manifold matrix has the form as ${\bf{A}}{\rm{ = }}\left[ {{\bf{a}}\left( {{\phi _1}} \right), \cdots ,{\bf{a}}\left( {{\phi _K}} \right)} \right]$, where  array steel vector ${\bf{a}}\left( {{\phi _k}} \right) = {\left[ {\exp \left( {j{\phi _k}0} \right), \cdots ,\exp \left( {j{\phi _k}\left( {M - 1} \right)} \right)} \right]^{\rm T}}$, and spatial phase
\begin{align}\label{Phi}
{\phi _k} = \frac{{\pi {d}\sin \left( {{\theta _k}} \right){f_k}}}{{{f_N}}},
\end{align}
where $d$ is the distance between two consecutive antennas in half-wavelengths corresponding to the Nyquist sampling rate ${f_N}$, ${\theta _k}$ and ${f_k}$ are the DOA and the center frequency of ${s_k}\left( t \right)$, respectively. The sensor position vector is ${\bf{d}} = \left[ {0,1, \cdots ,M - 1} \right]d$.  Note that the array is not limited to ULA for our receiver architecture and algorithms in the next few sections.}

The frequency domain output can be written as
\begin{align}\label{eqnDOAfre}
{\bf{X}}\left( f \right) = {\bf{AS}}\left( f \right) + {\bf{N}}\left( f \right),
\end{align}
where ${\bf{X}}\left( f \right) = {\left[ {{X_1}\left( f \right), \cdots ,{X_M}\left( f \right)} \right]^{\rm{T}}}$, ${\bf{S}}\left( f \right) = {\left[ {{S_1}\left( f \right), \cdots ,{S_K}\left( f \right)} \right]^{\rm{T}}}$, and ${\bf{N}}\left( f \right) = {\left[ {{N_1}\left( f \right), \cdots ,{N_M}\left( f \right)} \right]^{\rm{T}}}$ are the frequency domain expression of ${\bf{x}}\left( t \right)$, ${\bf{s}}\left( t \right)$, ${\bf{n}}\left( t \right)$, respectively. ${X_m}\left( f \right)$ is the Fourier transform of ${x_m}\left( t \right)$.
\subsection{ Objective statement}
The objective of this paper is to simultaneously estimate the carrier frequency ${f_k}$ and DOA ${\theta _k}$ of multiple signals ${s_k(t)}$. To achieve this goal, we will introduce the novel methods under the Nyquist sampling rate as follows.

\section{Proposed receiver architecture and signal reception model}
Now, we modify the traditional array signal receiver architecture and introduce the sub-Nyquist sampling technique into the architecture to reduce sampling rate. In this section, a novel architecture is presented and the corresponding signal reception model is derived.

\subsection{Receiver architecture}
\begin{figure}[!t]
\centering
\includegraphics[width=2.5in]{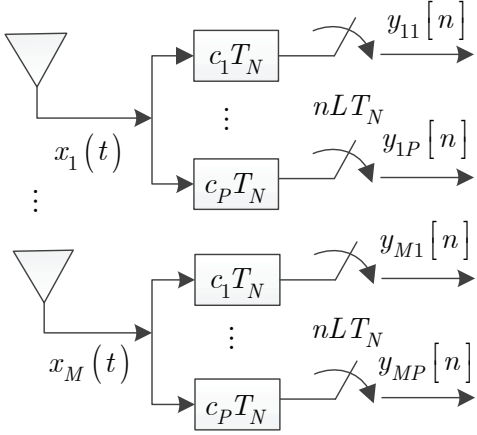}
\caption{Proposed receiver architecture.}
\label{figArcFull}
\end{figure}

Our receiver architecture is shown in Fig.\ref{figArcFull}. We apply the multi-coset sampling \cite{Mishali2009} in Fig.\ref{figArcFull} as representative of sub-Nyquist sampling technology.
In Fig. \ref{figArcFull}, there are $M$ sensors and every sensor is followed by $P$ delay branches. All the ADCs are well-synchronized and sample at a sub-Nyquist sampling rate of ${f_{sub}} = {{{f_N}} / L}$, where ${f_N} = {1 / {{T_N}}}$ is the Nyquist sampling rate  and $L$ is the sampling rate reduction factor, { where ${T_N}$ is Nyquist sampling interval}. The constant set $C=[c_1,\cdots,c_P]$ is referred to the sampling pattern where $0 \le {c_1} < {c_2} <  \cdots  < {c_P} \le L - 1$. ${y_{mp}}\left[ n \right]$ denotes the sampled signal corresponding to the $m$th sensor, $p$th branch.

The  average sampling rate of the multi-coset sampling is
 \begin{align}\label{feq}
{f_E} = \frac{{P{f_N}}}{L},
\end{align}
which is lower than the Nyquist rate $f_N$ when $P<L$.

\subsection{Signal reception model}
According to the conclusion of \cite{Mishali2009}, the relationship between the discrete-time Fourier transform ${Y_{mp}}\left( {{e^{j2\pi fT_N}}} \right)$ of the signal ${y_{mp}}\left[ n \right]$ and the Fourier transform ${X_m}\left( f \right)$ of ${x_m}\left( t \right)$ is as follows.
 \begin{align}\label{eqnSubNyVec}
{Y_{mp}}\left( {{e^{j2\pi f{T_N}}}} \right)&=\frac{1}{{L{T_N}}}\sum\limits_{l = 1}^L {\exp \left( {j\frac{{2\pi }}{L}{c_p}l} \right){X_{ml}}\left( f \right)} ,\nonumber\\
 & 0 \le p \le P,f \in \mathcal{F} \triangleq \left[ {0,f_{sub}} \right).
 \end{align}
The matrix form of (\ref{eqnSubNyVec}) is expressed as
\begin{align}\label{eqnSubNy}
{{\bf{Y}}_m}\left( f \right) = {\bf{B}}{\overline {\bf{X}} _m}\left( f \right),f \in \mathcal{F}, 1 \leq m \leq M,
\end{align}
where ${{\bf{Y}}_m}\left( f \right) = {\left[ {{Y_{m1}}\left( {{e^{j2\pi fT_N}}} \right), \cdots ,{Y_{mP}}\left( {{e^{j2\pi fT_N}}} \right)} \right]^{\rm{T}}}$,
${\overline {\bf{X}} _m}\left( f \right) = {\left[ {{X_{m1}}\left( f \right), \cdots ,{X_{mL}}\left( f \right)} \right]^{\rm{T}}}$,
${X_{ml}}\left( f \right) = {X_m}\left( {f + \left(l-1\right){f_{sub}}} \right)$,
${{{B}}_{il}} = \frac{1}{{LT_N}}\exp \left( {j\frac{{2\pi }}{L}{c_i}l} \right)$.
For convenience,  we multiply both sides of (\ref{eqnSubNy}) by $\sqrt L T_N$ to normalize the row vectors of $\bf{B}$. Then redefine ${B_{il}} \triangleq \frac{1}{{\sqrt L }}\exp \left( {j\frac{{2\pi }}{L}{c_i}l} \right)$, ${Y_{mp}}\left( f \right) \triangleq \sqrt L T_N{Y_{mp}}\left( {{e^{j2\pi fT_N}}} \right)$.

\noindent From (\ref{eqnDOAfre}),
\begin{align}\label{eqnunionEl}
{X_{ml}}\left( f \right) = {{\bf{A}}_m}{\widehat {\bf{S}}_l}\left( f \right)+{N_{ml}}\left( f \right),1 \le l \le L,f \in \mathcal{F},
\end{align}
where ${\widehat {\bf{S}}_l}\left( f \right) = {\left[ {{S_{1l}}\left( f \right), \cdots ,{S_{Kl}}\left( f \right)} \right]^{\rm{T}}}$, ${S_{kl}}\left( f \right) = {S_k}\left( {f + \left(l-1\right){f_{sub}}} \right)$, ${N_{kl}}\left( f \right) = {N_k}\left( {f + \left(l-1\right){f_{sub}}} \right)$.

\noindent In matrix form, (\ref{eqnunionEl}) can be expressed as
\begin{align}\label{eqnArrayFreSub}
{\overline {\bf{X}} _m}\left( f \right) = \left({{{\bf{I}}_L} \otimes {{\bf{A}}_m}} \right) \widehat {\bf{S}}\left( f \right)+ \widehat {\bf{N}}_m\left( f \right), f \in \mathcal{F},
\end{align}
where $\widehat {\bf{S}}\left( f \right) = {\left[ {\widehat {\bf{S}}_1^{\rm{T}}\left( f \right), \cdots ,\widehat {\bf{S}}_L^{\rm{T}}\left( f \right)} \right]^{\rm{T}}}$, ${\widehat {\bf{N}} }_m\left( f \right) = \left[ {{N_{m1}}\left( f \right), \cdots ,{N_{ml}}\left( f \right)} \right]^{\rm T}$.

\noindent Substituting (\ref{eqnArrayFreSub}) into (\ref{eqnSubNy}), we get
\begin{align}\label{eqnOneCh}
{{\bf{Y}}_m}\left( f \right) &= {\bf{B}}\left({{{\bf{I}}_L} \otimes {{\bf{A}}_m}} \right)\widehat {\bf{S}}\left( f \right) + {\bf{B}}{\widehat {\bf{N}}_m}\left( f \right) \nonumber \\
 &=\left( {{{\bf{A}}_m} \otimes {\bf{B}}} \right)\overline {\bf{S}} \left( f \right)  + {\bf{B}}{\widehat {\bf{N}}_m}\left( f \right),f \in \mathcal{F}, 1 \leq m \leq M,
\end{align}
where $\overline {\bf{S}} \left( f \right) = {\left[ {\overline {\bf{S}} _1^{\rm{T}}\left( f \right), \cdots ,\overline {\bf{S}} _K^{\rm{T}}\left( f \right)} \right]^{\rm{T}}}$, ${\overline {\bf{S}} _k}\left( f \right) = {\left[ {{S_{k1}}\left( f \right), \cdots ,{S_{kL}}\left( f \right)} \right]^{\rm{T}}}$.

\noindent Then, combining all $m$ can result in
\begin{align}\label{Yf}
{\bf{Y}}\left( f \right) &= \left( {{\bf{A}} \otimes {\bf{B}}} \right)\overline {\bf{S}} \left( f \right)  + \left( {{{\bf{I}}_M} \otimes {\bf{B}}} \right)\widehat {\bf{N}}\left( f \right)\\
 &\buildrel \Delta \over= {\bf{G}}\overline {\bf{S}} \left( f \right)+{{\bf{I}}_{\bf{B}}}\widehat {\bf{N}}\left( f \right), f \in \mathcal{F},\label{Yf2}
\end{align}
where ${\bf{Y}}\left( f \right) = {\left[ {{\bf{Y}}_1^{\rm{T}}\left( f \right), \cdots ,{\bf{Y}}_M^{\rm{T}}\left( f \right)} \right]^{\rm{T}}}$, $\widehat {\bf{N}}\left( f \right) = {\left[ {\widehat {\bf{N}}_1^{\rm{T}}\left( f \right), \cdots ,\widehat {\bf{N}}_M^{\rm{T}}\left( f \right)} \right]^{\rm{T}}}$. Actually, ${\bf{Y}}\left( f \right)$ in (\ref{Yf}) is the  matrix form of  the output of all branches of all sensors.

Because ${s_k}\left( t \right)$ is a narrowband signal, there is only one frequency band which is occupied in ${\overline{{\bf{S}}}_k}\left( f \right)$. Further, ${\overline{{\bf{S}}}_{k}}\left( f \right)$ is a sparse vector of length $L$ when $k$ is fixed and there is only one index ( marked as ${l_k}$), which is activated. Since it is assumed that those signals' carrier frequencies are far between, any two signals are not in the same sub-band. Namely, ${l_i}$ is not equal to ${l_j}$ for any $i \neq j$.  $\Omega  = \left[ {{l_1}, \cdots ,{l_K}} \right]$ denotes  the activated index set of ${\overline{{\bf{S}}}_{k}}\left( f \right)$ and $\bf{B}$. Further, $\overline {\bf{S}} \left( f \right)$ is a $K$-sparse vector of length $KL$. The support index $\mathcal{S}$ of $\overline {\bf{S}} \left( f \right)$ and ${\bf{G}}$ is determined as
\begin{align}\label{Supp}
\mathcal{S}_k = \left( {k - 1} \right)L + {l_k}.
\end{align}
With the knowledge of $\mathcal{S}$, (\ref{Yf2}) can be written as
\begin{align}\label{SparMod}
{\bf{Y}}\left( f \right)&= {{\bf{G}}_{\cal S}}{\overline {\bf{S}} ^{\cal S}}\left( f \right)+{{\bf{I}}_{\bf{B}}}\widehat {\bf{N}}\left( f \right)\\
&=\left( {{\bf{A}} * {{\bf{B}}_\Omega }} \right){\overline {\bf{S}} ^ {\cal S}}\left( f \right)+{{\bf{I}}_{\bf{B}}}\widehat {\bf{N}}\left( f \right),f \in \mathcal{F}.
\end{align}

\begin{remark}
{It is clear that (\ref{eqnSubNy}) and (\ref{eqnunionEl}) are the sub-Nyquist sampling model and DOA model, respectively. If we just severally consider the frequency estimation and DOA estimation in  (\ref{eqnSubNy}) and (\ref{eqnunionEl}), we will meet the match problem and can not comprehensively classify the targets. The target classification can be jointly handled in a union space based on equation (\ref{SparMod}), where the compressive sampling can be applied in time and space domain, respectively. As shown in Fig.\ref{fig2DSpace} and explained in section I, we have more chances and better performance to classify the targets in the time-space union domain. In terms of the performance, we will further study in Section V.  On the contrary, \cite{Ariananda2013} gives the 2D power spectrum instead of reception model. In \cite{Kumar2014, Kumar2015}, the sub-Nyquist sampling is also applied to array receiver, but only separate  models are given. Besides, those methods in~\cite{Kumar2015,Kumar2016}  are limited to ULA because they make use of the rotation invariance property of ULA. Their receiver can not employ other particular array to make use of those arrays' advantages.}
\end{remark}

In (\ref{Yf}), since the Khatri-Rao product is used to unify the frequency domain and spatial domain into a two-dimensional matrix form, this equation can be viewed from the perspective of third-order tensor. We will discuss the third-order tensor in the next section. On the other hand, if  ${\bf{G}}$  is regarded as  a special array manifold, the subspace decomposition theory can be employed. Of course, the different perspectives will derive different methods, which will be analyzed in detail.

\section{Joint DOA and Frequency Estimation Algorithm}
\subsection{ Algorithm based on trilinear decomposition}
It is easy to express (\ref{SparMod}) in element form as
\begin{align}\label{TriDec}
{Y_{mp}}\left( f \right)=\sum\limits_{k = 1}^K {{{{A}}_{mk}}{{{B}}_{p{l_k}}}{{{S}}_{k{l_k}}}\left( f \right)},f \in \mathcal{F}.
\end{align}
From (\ref{TriDec}), ${Y_{mp}}\left( f \right)$  can be regarded as a third-order tensor and the problem can be solved by trilinear decomposition \cite{Kruskal1977, Lathauwer2006}. This problem is different from the standard trilinear decomposition problem since ${\bf{B}}$ is known here. Even so, we do not know which columns are activated. So, we can use the standard trilinear decomposition algorithm, such as alternating least squares (ALS) \cite{DeLathauwer2008} and regularized alternating least squares (RALS) \cite{Li2013,Navasca2008}, where some sufficient conditions for uniqueness up to permutation and scalings of the decomposition are provided. After the decomposition, we can obtain $\widetilde {\bf{A}}$, ${\widetilde {\bf{B}}}$, and $\widetilde {\bf{S}}\left( f \right)$.

Since not only every column vector of $\widetilde {\bf{A}}$ but also every row vector of  $\widetilde {\bf{s}}\left( t \right)$ (the inverse discrete-time Fourier transform of $\widetilde {\bf{S}}\left( f \right)$) can be viewed as a single tone, {periodogram is applied} on every column vector of $\widetilde {\bf{A}}$ or every row vector of  $\widetilde {\bf{s}}\left( t \right)$  to achieve $\phi$ or $f$  {maximum likelihood (ML) } estimation \cite{Rife1974}. The periodogram is briefly introduced in the following. {For an $N$-length single frequency sine wave $z\left( n \right) = \exp \left( {j\omega_0 n} \right),n = 1,2, \cdots ,N$, the ML estimation for $\omega_0$ is realized through}
\begin{align}\label{MLSF}
{\hat \omega _0  = \arg \mathop {\max }\limits_{\omega  \in \left[ {0,2\pi } \right)} \left| {\sum\limits_{n = 1}^N {z\left( n \right)} \exp \left( { - j\omega n} \right)} \right|}.
\end{align}
{Similarly, for an arbitrary array form ${\bf{z}}={\bf{a}}\left( {{\phi}_0} \right)$, the ML estimation for $\phi_0$ is realized by}
\begin{align}\label{MLSFAny}
{\hat \phi_0  = \arg \mathop {\max }\limits_{\phi  \in \left[ {0,2\pi } \right)} \left| {{{\bf{a}}^{\rm{H}}}\left( \phi  \right){\bf{z}}} \right|.}
\end{align}

We determine $\Omega$ through comparing the correlation coefficient of the column between $\bf{B}$ and $\widetilde{{\bf{B}}}$ as
\begin{align}\label{Omega}
{\Omega _k} = \mathop {\arg \max }\limits_j {r_{kj}} = \left| {\frac{{{{\left( {{{\widetilde {\bf{B}}}^k}} \right)}^{\rm{H}}}{{\bf{B}}^j}}}{{\left\| {{{\widetilde {\bf{B}}}^k}} \right\|\left\| {{{\bf{B}}^j}} \right\|}}} \right|,j = 1, \cdots ,L.
\end{align}

The received signal's frequency estimation ${\overline f _k}$  is obtained by applying periodogram to $\widetilde {\bf{s}}_k\left( t \right)$. Besides, there is a relationship between ${\overline f _k}$ and the original signal's frequency ${ f _k}$:
\begin{align}\label{MatchFreDir}
{f_k} = \left( {{\Omega _k} - 1} \right)\frac{{{f_N}}}{L} + \overline {{f}}_k.
\end{align}

And then, once $\phi_k$ and $f_k$ are known, $\theta_k $ can be acquired by (\ref{Phi}). We outline the main steps of this method named joint algorithm  based on trilinear decomposition (JDFTD) in table \ref{Alg1}.

\begin{table}[!t]
    \renewcommand{\arraystretch}{1.0}
    \caption{\textbf{Algorithm JDFTD}}\label{Alg1}
    \centering
    \begin{tabularx}{8.4cm}{lX}
        \toprule
         1)&Obtain $\widetilde {\bf{A}}$, ${\widetilde {\bf{B}}}$, and $\widetilde {\bf{S}}\left( f \right)$ using RALS according to (\ref{TriDec});\\
         2)&Gain $\phi_k$ applying (\ref{MLSF}) or (\ref{MLSFAny}) to $\bf{A}$;\\
         3)&Determine $\Omega_k $ according to (\ref{Omega});\\
         4)&Get $\widetilde {\bf{s}}_k\left( t \right)$ and ${\overline f _k}$ by applying IFFT, (\ref{MLSF}) to $\widetilde {\bf{S}}_k\left( f \right)$, $\widetilde {\bf{s}}_k\left( t \right)$, consecutively;\\
         5)&Compute ${ f _k}$ through (\ref{MatchFreDir});\\
         6)&Calculate $\theta_k$ through (\ref{Phi});\\
        \bottomrule
    \end{tabularx}
\end{table}

\subsection{Algorithm based on  subspace decomposition}
In this subsection, we will take advantage of subspace decomposition theory \cite{Schmidt1986}. The covariance matrix of ${\bf{Y}}\left( f \right)$, $f \in \mathcal{F}$ is given by
\begin{align}\label{eqtR}
{\bf{R}} = {\mathop{\rm E}\nolimits} \left( {{\bf{Y}}\left( f \right){{\bf{Y}}^{\rm H}}\left( f \right)} \right){\rm{ = }}{{\bf{G}}_S}{{\bf{R}}_{\overline {\bf{S}} }}{\bf{G}}_S^{\rm{H}} + {\sigma ^2}{{\bf{I}}_{MP}},
\end{align}
where ${{\bf{R}}_{\overline {\bf{S}} }}$ , ${\sigma ^2}{{\bf{I}}_{MP}}$ are the source and noise covariance matrix, respectively. (\ref{eqtR}) makes use of ${{\bf{I}}_{MP}} = {{\bf{I}}_{\bf{B}}}{\bf{I}}_{\bf{B}}^{\rm H}$. In actual situation, we can obtain the estimate of the autocovariance matrix through
\begin{align}\label{eqtRh}
{\bf{R}} = \frac{1}{{T/L}}\sum_{f = 1}^{T/L} {{\bf{Y}}\left( f \right){{\bf{Y}}^{\rm H}}\left( f \right)}
\end{align}
when {$T$} the snapshots of observation are sufficient.

Applying the singular value decomposition (SVD) to $\bf{R}$ results in
\begin{align}\label{SVDR}
{\bf{R}}= {{\bf{U}}_S}{{\bf{D}}_S}{\bf{U}}_S^{\rm H} + {{\bf{U}}_N}{{\bf{D}}_N}{\bf{U}}_N^{\rm H},
\end{align}
where ${{\bf{U}}_S}$ and ${{\bf{U}}_N}$ are  signal subspace and noise subspace, respectively. Since the signal subspace and the noise subspace are orthogonal,
\begin{align}\label{Orth}
{{\bf{a}}_{l}}\left( \phi  \right) \bot {{\bf{U}}_N}
\end{align}
holds, where ${{\bf{a}}_l}\left( \phi  \right) = {\bf{a}}\left( \phi  \right) \ast {{\bf{B}}_l}$. Computing the pseudo-spectra
\begin{align}\label{MUSICP}
P\left( {l,\phi } \right) = \frac{1}{{{{\left\| {{\bf{a}}_l^{\rm{H}}\left( \phi  \right){{\bf{U}}_N}} \right\|}^2}}}
\end{align}
and applying a peak search algorithm, $\phi_k$, $l_k$ are obtained. Further, we have ${\bf{A}}$, $\Omega$. Since ${\mathcal{S}}$ can be solved by (\ref{Supp}), the least square solution of ${\overline {\bf{S}} ^\mathcal{S} }\left( f \right)$ is given by
\begin{align}\label{MLES}
\overline{\bf{S}}^{\mathcal{S}}\left( f \right)={\bf{G}}_\mathcal{S} ^\dag {\bf{Y}}\left( f \right), f \in \mathcal{F},
\end{align}
where ${{\bf{G}}_{\mathcal{S}}} = {\bf{A}}*{{\bf{B}}_\Omega }$. Similarly, $f_k$ and $\theta_k$ can be calculated through the step 4)-6) of Table \ref{Alg1}. We outline the main steps of this method named joint algorithm based on subspace decomposition (JDFSD) in table \ref{Alg2}.

\begin{table}[!t]
    \renewcommand{\arraystretch}{1.0}
    \caption{\textbf{Algorithm JDFSD}}\label{Alg2}
    \centering
    \begin{tabularx}{8.4cm}{lX}
        \toprule
         1)&Calculate ${\bf{R}}$ according to (\ref{eqtRh});\\
         2)&Gain ${{\bf{U}}_N} $ by applying SVD to  ${\bf{R}}$;\\
         3)&Compute $P\left(l, \phi  \right)$  through (\ref{MUSICP});\\
         4)&Acquire $\phi_k$, $l_k$ by peak search algorithm, further, we have ${\bf{A}}$, $\Omega$, and ${{\bf{G}}_{\mathcal{S}}}$;\\
         5)&Determine $\overline{\bf{S} }^{\mathcal{S}}\left( f \right)$ according to (\ref{MLES});\\
         6)&Execute the step 4)-6) of Table \ref{Alg1}.\\
        \bottomrule
    \end{tabularx}
\end{table}

\section{Cram\'{e}r\text{-}Rao Bound} \label{SectionCRB}
In this section, we derive the CRB on the covariance matrix of any unbiased estimator of $\phi$.
If the signal autocorrelation matrices are defined as
\begin{align}\label{DRs}
&{\bf{R}}_{\bf{S}} \buildrel\Delta \over = \frac{1}{T}\sum_{f = 1}^T {{\bf{S}}\left( f \right){\bf{S}}{{\left( f \right)}^{\rm{H}}}} \\
&{{\bf{R}}_{\overline {\bf{S}} }} \buildrel \Delta \over = \frac{1}{T/L }\sum_{f = 1}^{T/L} {{{\overline {\bf{S}} }^{\cal S}}\left( f \right){{\overline {\bf{S}} }^{\cal S}}^{\rm{H}}\left( f \right)}
\end{align}
and according to the exchangeability of summing and the definition of  ${{\overline {\bf{S}} }^{\mathcal{S}}}\left( f \right)$,
\begin{align}\label{TwoRs}
{\bf{R}}_{{\bf{S}}}   = \frac{1}{L} {\bf{R}}_{\overline  {\bf{S}}}
\end{align}
holds.

{Based on} the form of ${\bf{B}}$, ${{\bf{R}}_{\bf{B}}} = {{\bf{I}}_{\bf{B}}}{{\bf{I}}_{\bf{B}}}^{\rm H} = {{\bf{I}}_{MP}}$ holds. According to model (\ref{Yf}), the log\text{-}likelihood function of the data ${\bf{Y}}\left( f \right)$, $f \in \mathcal{F}$ is given by
\begin{align}\label{logLF}
\ln L = \;&const  - \frac{1}{\sigma ^2}\sum_{f = 1}^{T/L} {{{\left( {{\bf{Y}}\left( f \right) - {{\bf{G}}_{\mathcal{S}}}{{\overline {\bf{S}} }^{\mathcal{S}}}\left( f \right)} \right)}^{\rm{H}}}} \\
  &{\bf{\cdot}} \left( {{\bf{Y}}\left( f \right) - {{\bf{G}}_{\mathcal{S}}}{{\overline {\bf{S}} }^{\mathcal{S}}}\left( f \right)} \right). \nonumber
\end{align}
Comparing  (\ref{logLF}) with APPENDIXE (E.1) in \cite{Stoica1990}, and making use of the conclusion of Section IV equation (4.6) in \cite{Stoica1990},  the CRB of our model is given by
\begin{align}\label{CRBsubP}
{\rm{CRB}}_{sub} &= \frac{\sigma ^2}{{2T/L}}{\left( {\Re \left( {\left( {{{\bf{E}}^{\rm{H}}}{{\bf{P}}_{{{\bf{G}}_{\mathcal{S}}}}}{\bf{E}}} \right) \odot {\bf{R}}_{\overline {\bf{S}} }^{\rm{H}}} \right)} \right)^{ - 1}}\nonumber \\
&= \frac{{{\sigma ^2}}}{{2T}}{\left( {\Re \left( {\left( {{{\bf{E}}^{\rm{H}}}{{\bf{P}}_{{{\bf{G}}_{\mathcal{S}}}}}{\bf{E}}} \right) \odot {\bf{R}_{\bf{S}}^{\rm{H}}}} \right)} \right)^{ - 1}}
\end{align}
where ${{\bf{P}}_{{{\bf{G}}_{\mathcal{S}}}}} = {\bf{I}} - {{\bf{G}}_{\mathcal{S}}}{\bf{G}}_{\mathcal{S}}^\dag $, where ${\bf{G}}_{\mathcal{S}}^\dag  = {\left( {{\bf{G}}_{\mathcal{S}}^{\rm{H}}{{\bf{G}}_{\mathcal{S}}}} \right)^{ - 1}}{\bf{G}}_{\mathcal{S}}^{\rm{H}}$, ${\bf{E}} = \left[ {{{\bf{E}}_1}, \cdots ,{{\bf{E}}_K}} \right]$, ${{\bf{E}}_i} = \frac{{d{{{\bf{G}}_{{\mathcal{S}}_i}}}}}{{d{\phi _i}}}$.
For comparing, the CRB  which employs the Nyquist sampling (marked as $\textrm{CRB}_{Ny}$) \cite{Stoica1990} is rewritten here.
\begin{align}\label{CRBNyP}
{\rm{CRB }}_{Ny}= \frac{{{\sigma ^2}}}{{2T}}{\left( {\Re \left( {\left( {{{\bf{D}}^{\rm{H}}}{{\bf{P}}_{{{\bf{A}}}}}{\bf{D}}} \right) \odot {\bf{R}_{\bf{S}}^{\rm{H}}}} \right)} \right)^{ - 1}}
\end{align}
where ${{\bf{P}}_{\bf{A}}} = {\bf{I}} - {\bf{A}}{{\bf{A}}^\dag }$, ${\bf{D}} = \left[ {{{\bf{D}}_1}, \cdots ,{{\bf{D}}_K}} \right]$, ${{\bf{D}}_i} = \frac{{d{{\bf{A}}_i}}}{{d{\phi _i}}}$.

Next, we will show that ${\rm{CRB}}_{sub}$ is lower than ${\rm{CRB}}_{Ny}$  when the branch number of our architecture is equal to the sampling rate reduction factor ($P=L$). At this point, on the one hand, the sub-Nyquist sampling and Nyquist sampling obtain {equal} snapshot in the same time, on the other hand, the received data by sub-Nyquist sampling can just be viewed as the rearrangement of the received data by Nyquist sampling. The proof will carry out in two steps: first, $\textrm{CRB}_{Ny}$ will increase with the number of source $K$; secondly, ${\rm{CRB}}_{sub}$ is not influenced by  the number of source, and the ${\rm{CRB}}_{sub}$ is equal to the minimum value of $\textrm{CRB}_{Ny}$.

For convenience, let us introduce the following notation: ${{\bf{A}}_ + }=\left[ {{\bf{A}},a} \right]$, ${{\bf{D}}_ + }=\left[ {{\bf{D}},d} \right]$, ${{\bf{T}}_ + } = {{\bf{D}}}_ +^{\rm H}\left( {{\bf{I}} - {{\bf{A}}_ + }{{\bf{A}}}_ +^\dag } \right){{\bf{D}}_ + }$, ${\bf{T}} = {{\bf{D}}^{\rm H}}\left( {{\bf{I}} - {\bf{A}}{{\bf{A}}^\dag }} \right){\bf{D}}$, where $a$ is the steer vector corresponding to the increased angle $\phi_{K+1}$, $d = \frac{{da}}{{d\phi_{K+1} }}$. $ {{\bf{R}}_{\rm{ + }}}=\left[ {\begin{array}{*{20}{c}}
{\bf{R_S}}&\mu \\
{{\mu ^{\rm H}}}&\nu
\end{array}} \right]$ is the  covariance matrix of all $K+1$ signals, where ${\bf{R_S}}$ is the covariance matrix of original $K$ signals, $\mu $ is cross-correlation vector between the increased signal and the original signals, $\nu $ is  average power of  the increased signal.  ${\left( {\bf{R}} \right)_K}$ denotes the $K$th order principal minor determinant of $ {\bf{R}} $.

Making use of the nested structure of ${{\bf{A}}_ + }$  and the matrix inversion lemma \cite{Horn1985},
 \begin{align}\label{ApApInv}
{{\bf{A}}_ + }{{\bf{A}}}_ +^\dag = {\bf{A}}{{\bf{A}}^\dag } + {\bf{U}},
\end{align}
holds after some matrix manipulations, where ${\bf{U}}=\frac{1}{{{a^{\rm H}}\left( {{\bf{I}} - {\bf{A}}{{\bf{A}}^\dag }} \right)a}}\left( {\left( {{\bf{I}} - {\bf{A}}{{\bf{A}}^\dag }} \right)a} \right){\left( {\left( {{\bf{I}} - {\bf{A}}{{\bf{A}}^\dag }} \right)a} \right)^{\rm H}} \succeq 0$ since $\left( {{\bf{I}} - {\bf{A}}{{\bf{A}}^\dag }} \right)$ is a projection matrix.

\noindent  Taking (\ref{ApApInv}) and the nested structure of ${{\bf{D}}_ + }$ into ${{\bf{T}}_ + }$ leads to
\begin{align}\label{Tp}
{\left( {{{\bf{T}}_ + }} \right)_K}={\bf{T}} - {{\bf{D}}^{\rm H}}{\bf{UD}}.
\end{align}
 It is easy to proof that ${{\bf{D}}^{\rm H}}{\bf{UD}} \succeq 0$  with ${\bf{U}} \succeq 0$. We thus hold
\begin{align}\label{TpK}
{\left( {{{\bf{T}}_ + }} \right)_K}={\bf{T}} - {{\bf{D}}^{\rm H}}{\bf{UD}} \preceq {\bf{T}}.
\end{align}

To proceed, we give a proposition.
 \begin{proposition}\label{Lemma1}
 For a Hermitian matrix ${{\bf{M}}_{\rm{ + }}} \in {C^{\left( {K + 1} \right)\left( {K + 1} \right)}}$, if ${{\bf{M}}_{\rm{ + }}} \succeq 0$, then ${\left( {{{\bf{M}}}_ +^{ - 1}} \right)_K} \succ {\left( {{{\left( {{{\bf{M}}_ + }} \right)}_K}} \right)^{ - 1}}$.
\end{proposition}

\begin{IEEEproof}
Since ${{\bf{M}}_ + } \succeq 0$ it can be decomposed as ${{\bf{M}}_ + } = {{\bf{H}}_ + }{{\bf{H}}}_ +^{\rm H}$, where ${{\bf{H}}_ + } \in {{\mathcal{C}}^{K \times K}}$ \cite{Horn1985}. Dividing ${{\bf{H}}_ + }$ into blocks as ${{\bf{H}}_ + } = \left[ {{\bf{H}},h} \right]$ results in
\begin{align}
{{\bf{M}}_ + } = \left[ {\begin{array}{*{20}{c}}
{{{\bf{H}}^{\rm H}}{\bf{H}}}&{{{\bf{H}}^{\rm H}}h}\\
{{h^{\rm H}}{\bf{H}}}&{{h^{\rm H}}h}
\end{array}} \right] = \left[ {\begin{array}{*{20}{c}}
{\bf{M}}&{{{\bf{H}}^{\rm H}}h}\\
{{{\bf{H}}^{\rm H}}h}&{{h^{\rm H}}h}
\end{array}} \right].
\end{align}
Making use of the matrix inversion lemma \cite{Horn1985},
\begin{align}\label{MppInv}
{\left( {{\bf{M}}_ + ^{ - 1}} \right)_K}={{\bf{M}}^{ - 1}}{\rm{ + }}{\bf{V}} \succeq {{\bf{M}}^{ - 1}}={\left( {{{\left( {{{\bf{M}}}_ +} \right)}_K}} \right)^{ - 1}}
\end{align}
holds  since ${\bf{V}}=\frac{1}{{{h^{\rm H}}\left( {{\bf{I}} - {\bf{H}}{{\bf{H}}^\dag }} \right)h}}{\left( {{{\bf{M}}^{ - 1}}{{\bf{H}}^{\rm H}}h} \right)^{\rm H}}\left( {{{\bf{M}}^{ - 1}}{{\bf{H}}^{\rm H}}h} \right) \succeq 0$.
\end{IEEEproof}

\noindent Note that ${{\bf{T}}_+}={{\bf{D}}}_ +^{\rm H}\left( {{\bf{I}} - {{\bf{A}}_ + }{{\bf{A}}}_ +^\dag } \right){{\bf{D}}_ + } \succeq 0$, ${{\bf{R}}_{\rm{ + }}} \succeq 0$, $\Re \left( {{{\bf{T}}_{\rm{ + }}} \odot {\bf{R}}_+^{\rm H}} \right) \succeq 0$ holds.  Making use of proposition \ref{Lemma1}, we then obtain
 \begin{align}\label{InvKey}
{\left( {{{\left( {\Re \left( {{{\bf{T}}_{\rm{ + }}} \odot {{\bf{R}}_{\rm{ + }}}} \right)} \right)}^{{\rm{ - }}1}}} \right)_K} &\succeq {\left( {\Re {{\left( {{{\bf{T}}_{\rm{ + }}} \odot {{\bf{R}}_{\rm{ + }}}} \right)}_K}} \right)^{{\rm{ - }}1}}\nonumber\\
&={\left( {\Re \left( {{{\left( {{{\bf{T}}_ + }} \right)}_K} \odot {\bf{R_S}}} \right)} \right)^{{\rm{ - }}1}}\nonumber\\
&\succeq {\left( {\Re \left( {{\bf{T}} \odot {\bf{R_S}}} \right)} \right)^{{\rm{ - }}1}}.
\end{align}

Considering $\textrm{CRB}_{ Ny+ } = \frac{{{\sigma ^2}}}{{2T}}{\left( {\Re \left( {{{\bf{T}}_{\rm{ + }}} \odot {\bf{R}}_+^{\rm H}} \right)} \right)^{{\rm{ - }}1}}$, $\textrm{CRB}_{ Ny } = \frac{{{\sigma ^2}}}{{2T}}{\left( {\Re \left( {{\bf{T}} \odot {{\bf{R}}_{\bf{S}}^{\rm H}}} \right)} \right)^{{\rm{ - 1}}}}$, we get
 \begin{align}\label{CRBadd}
{\left( {\textrm{CRB}_{ Ny+ }} \right)_K} &= \frac{{{\sigma ^2}}}{{2T}}{\left( {{{\left( {\Re \left( {{{\bf{T}}_{\rm{ + }}} \odot {\bf{R}}_{\rm{ + }}^{\rm H}} \right)} \right)}^{{\rm{ - }}1}}} \right)_K} \nonumber \\
 &\succeq \frac{{{\sigma ^2}}}{{2T}}{\left( {\Re \left( {{\bf{T}} \odot {{\bf{R}}_{\bf{S}}^{\rm H}}} \right)} \right)^{{\rm{ - 1}}}} = \textrm{CRB}_{ Ny }.
\end{align}
(\ref{CRBadd}) shows that the estimate performance for $\bm{\phi}  = \left[ {{\phi _1}, \cdots ,{\phi _K}} \right]$ in the scene where there are only $K$ signals from $\bm{\phi}$ is better than that in the scene where there are both the $K$ signals from $\bm{\phi}$ and the increased signal from $\phi_{K+1}$. In other words, the increase of the number of DOA will degrade the performance of DOA estimate. It is simplistic to conclude that the estimation variance is lowest when there is only one signal. After calculation, the lowest estimation variance is $\frac{6}{{{\rm{SNR}} \cdot T \cdot M\left( {{M^2} - 1} \right)}}$, {where {signal-to-noise ratio (SNR)} is defined as  ${\rm{SN}}{{\rm{R}}} = (E({\left| {{{\bf{s}}}(t)} \right|^2})/{\sigma ^2})$}.

However, the performance of DOA estimation based on the proposed model will not degrade with the increase of the number of DOAs. It's easy to get ${\bf{E}} = {\bf{D}} \ast {\bf{B}}_\Omega$ since ${{\bf{G}}_{\mathcal{S}}} = {\bf{A}} \ast {\bf{B}}_\Omega$. We further hold that
 \begin{align}\label{SteerCor}
{{\bf{G}}}_{\mathcal{S}}^{\rm H}{{\bf{G}}_{\mathcal{S}}} &= \left( {{{\bf{A}}^{\rm H}}{\bf{A}}} \right) \odot \left( {{{\bf{B}}}_\Omega^{\rm H}{{\bf{B}}_\Omega }} \right),\\
{{\bf{E}}^{\rm H}}{\bf{E}} &= \left( {{{\bf{D}}^{\rm H}}{\bf{D}}} \right) \odot \left( {{{\bf{B}}}_\Omega^{\rm H}{{\bf{B}}_\Omega }} \right).
\end{align}
Thus, ${{\bf{G}}}_{\mathcal{S}}^{\rm H}{{\bf{G}}_{\mathcal{S}}} = \left( {{{\bf{A}}^{\rm H}}{\bf{A}}} \right) \odot {\bf{I}}$, ${{\bf{E}}^{\rm H}}{\bf{E}} = \left( {{{\bf{A}}^{\rm H}}{\bf{A}}} \right) \odot {\bf{I}}$ hold when $P=L$. Based on above results and after some matrix manipulations, we know that the the estimation variance based on our model  maintains  at $\frac{6}{{{\rm{SNR}} \cdot T \cdot M\left( {{M^2} - 1} \right)}}$ all the time when $P=L$. Consequently, when $L=P$, we have
\begin{align}\label{CRBK}
\begin{array}{l}
{\rm{CR}}{{\rm{B}}_{sub}} = {\rm{CR}}{{\rm{B}}_{Ny}}, \text{when} \: K = 1,\\
{\rm{CR}}{{\rm{B}}_{sub}} \le {\rm{CR}}{{\rm{B}}_{Ny}}, \text{when} \: K > 1.
\end{array}
\end{align}

\begin{figure}[!t]
\centering
\includegraphics[width=3in]{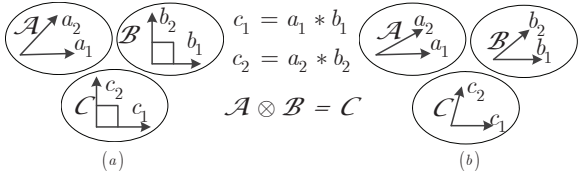}
\caption{2D union space.}
\label{fig2DSpace}
\end{figure}

\begin{remark}
We can view (\ref{CRBK}) from the physical perspective. {If we study the cross-correlation $\delta_{ij}$ between ${\bf{B}}^i$ and ${\bf{B}}^j$, i.e. $\delta_{ij} = \left| {{{\left( {{{\bf{B}}^i}} \right)}^{\rm{H}}}{{\bf{B}}^j}} \right|$,$i \neq j$, it will be easy to obtain}
 \begin{align}\label{CorB}
\left\{ {\begin{array}{*{20}{c}}
{{\delta _{ij}} < 1,P < L}\\
{{\delta _{ij}} = 0,P = L}
\end{array}} \right..
\end{align}
Since (\ref{SteerCor}) and (\ref{CorB}) hold, the cross-correlation of the new steer vectors ${{\bf{G}}_{\mathcal{S}}}$  is lower than that of the primary steer vectors ${\bf{A}}$ no matter whether $P$ is equal to $L$ or not.
Specifically, the new steer vectors are completely uncorrelated when $P=L$ in spite of the primary steer vectors are correlated. At this time, the new DOA estimation is equivalent to execution one by one in a scene where there is only one signal. This is an explanation why the performance of DOA estimation based on our model will not degrade with the increasing of the DOA number  while the performance of DOA estimation based on the primary model will degrade.
{
From the perspective geometry, this situation is corresponding to  Fig.\ref{fig2DSpace} (a), where as long as the vectors in space $\mathcal{B}$  are orthogonal, vectors in space $\mathcal{C}$  will be orthogonal no matter whether the vectors in space $\mathcal{A}$ are orthogonal, even if they are same. A more general case i.e. $P<L$, is corresponding to  Fig.\ref{fig2DSpace} (b), where even if the vectors in space neither $\mathcal{A}$ nor $\mathcal{B}$ are orthogonal, vectors in space $\mathcal{C}$  will be less correlated,  i.e., the angle between the vectors in space $\mathcal{C}$ is larger than before. } In an subjective sense, the cross-correlation of steer vectors reflects similarity and identifiability of the DOA. The lower cross-correlation of  steer vectors, the easier they are to be distinguished.
\end{remark}

\section{Simulation}
In this section, we present some numerical simulations to illustrate the performance of the proposed algorithms. In our examples, we consider some complex-valued narrowband far-field non-coherent signals with equal power impinging on a ULA composed of $M=8$ sensors which are separated by a half wavelength corresponding to Nyquist sampling rate, which would probably be the signal highest frequency. In the simulations,  we fix the number of snapshots at $T = 4000$ for Nyquist sampling, $T_{sub} = T/L$ for sub-Nyquist sampling, the Nyquist sampling rate at $f_N=10$ GHz, and the sampling rate reduction factor at $L=20$.

The {root-mean-square error (RMSE)} of DOA is defined as ${\rm{RMSE}} = \sqrt {\frac{1}{{{N_m}K}}\sum\nolimits_{i = 1}^{{N_m}} {\sum\nolimits_{k = 1}^K {{{(\theta _k^i - \widehat \theta _k^i)}^2}} } } $,  where the superscript $i$ refers to the $i$th trial, ${N_m}$ denotes the number of Monte Carlo tests. And the definitions of RMSE of spatial phase and frequency are similar to that of DOA.

Later on, we will study the performance versus different noise levels, different branch number, or different source number. {We will compare our methods with ST-Euler-ESPRIT in~\cite{Kumar2016}. The receiver configuration parameters of ST-Euler-ESPRIT are the same with ours. The delay is Nyquist sampling interval $T_N=1/f_N$. Hereon, we give the reasons why we choose ST-Euler-ESPRIT: i) In ~\cite{Ariananda2013}, both  frequency and DOA estimation accuracy are low since they are limited by the reciprocal of block length and the array aperture, respectively. ii) In terms of the hardware complexity, \cite{Kumar2015} and ~\cite{Kumar2014}  is the simplified version \cite{Kumar2016}, and \cite{Kumar2016} has the same hardware complexity with our receiver. For the sake of fair, we compare our methods with \cite{Kumar2016}.}
 20000 Monte Carlo trials for each example are implemented in this section.

\subsection{Performance with noise}\label{SimuB}
In this subsection, it will be shown that our model can be solved by the proposed algorithm in different noise levels.  We add an array construction MRA to prove the validity of the algorithm in this subsection. The MRA is composed of $M=8$ sensors which are located at ${\bf{d}} = \left[ {0,1,4,10,16,22,28,30} \right]d$. {However} ST-Euler-ESPRIT will be feasible only when  ULA is employed.

In this subsection, we set the branch number $P=L$. Meanwhile, the average sampling rate ${f_A}=\frac{{P{f_N}}}{L} $ is equal to the Nyquist sampling rate. We consider that the signal number $K=3$, and signals are from ${\bm{\theta }} = \left[ {{\theta _1},{\theta _2},{\theta _3}} \right]$, where ${\theta _1}$, ${\theta _2}$, ${\theta _3}$  are subject to $[-12.5^\circ,-7.5^\circ]$, $[-2.5^\circ,2.5^\circ]$, $[7.5^\circ,12.5^\circ]$ uniformly distribution, respectively. And the signal carrier frequency ${\bf{f}} = \left[ {{f_1},{f_2},{f_3}} \right]$ are subject to $[0.5,9.5]$ GHz uniformly distribution, and  any two signals are not in the same sub-band.

\begin{figure}[!t]
\centering
\includegraphics[width=2.5in]{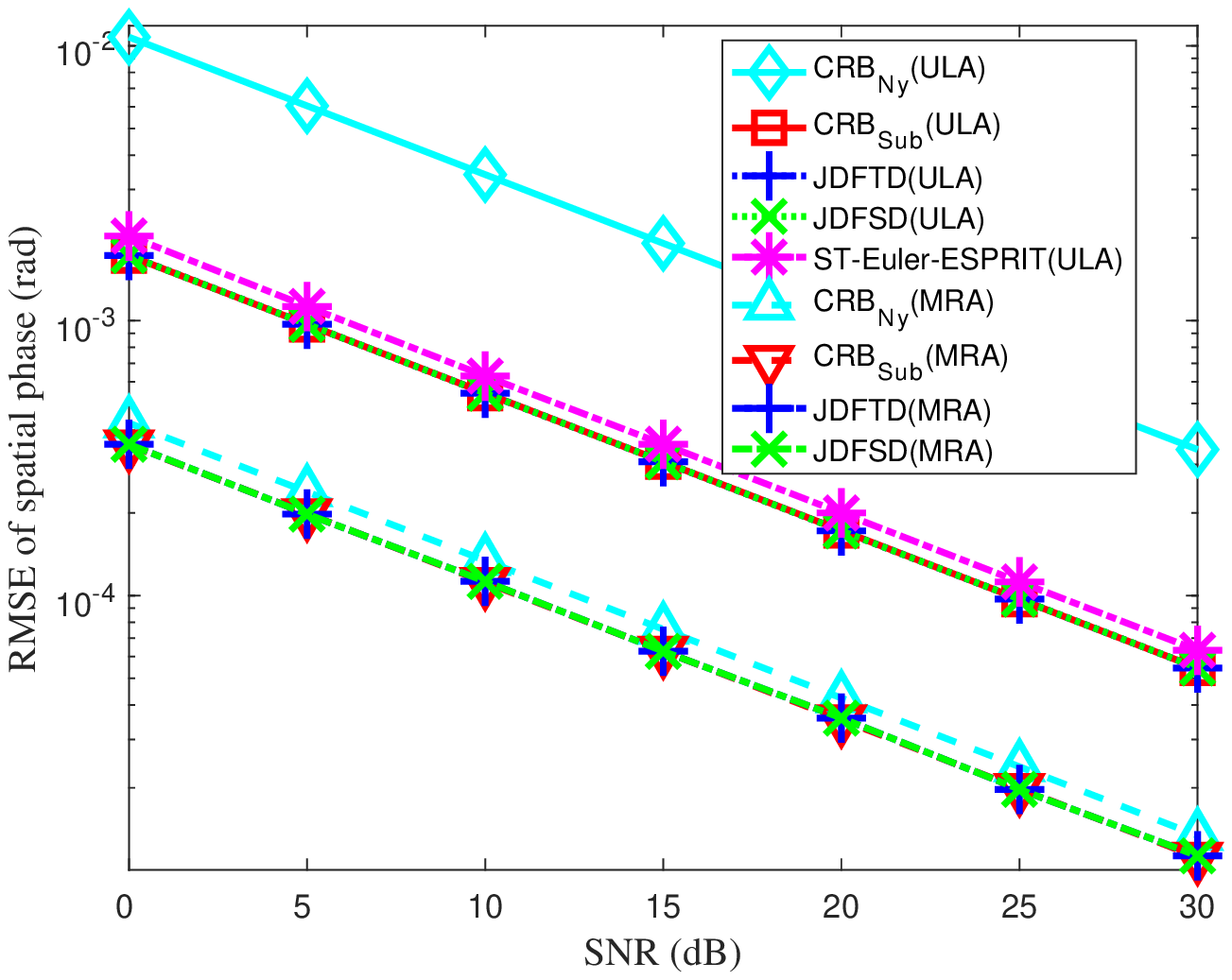}
\caption{RMSE of phase estimates versus SNR.}
\label{figPS}
\end{figure}

\begin{figure}[!t]
\centering
\includegraphics[width=2.5in]{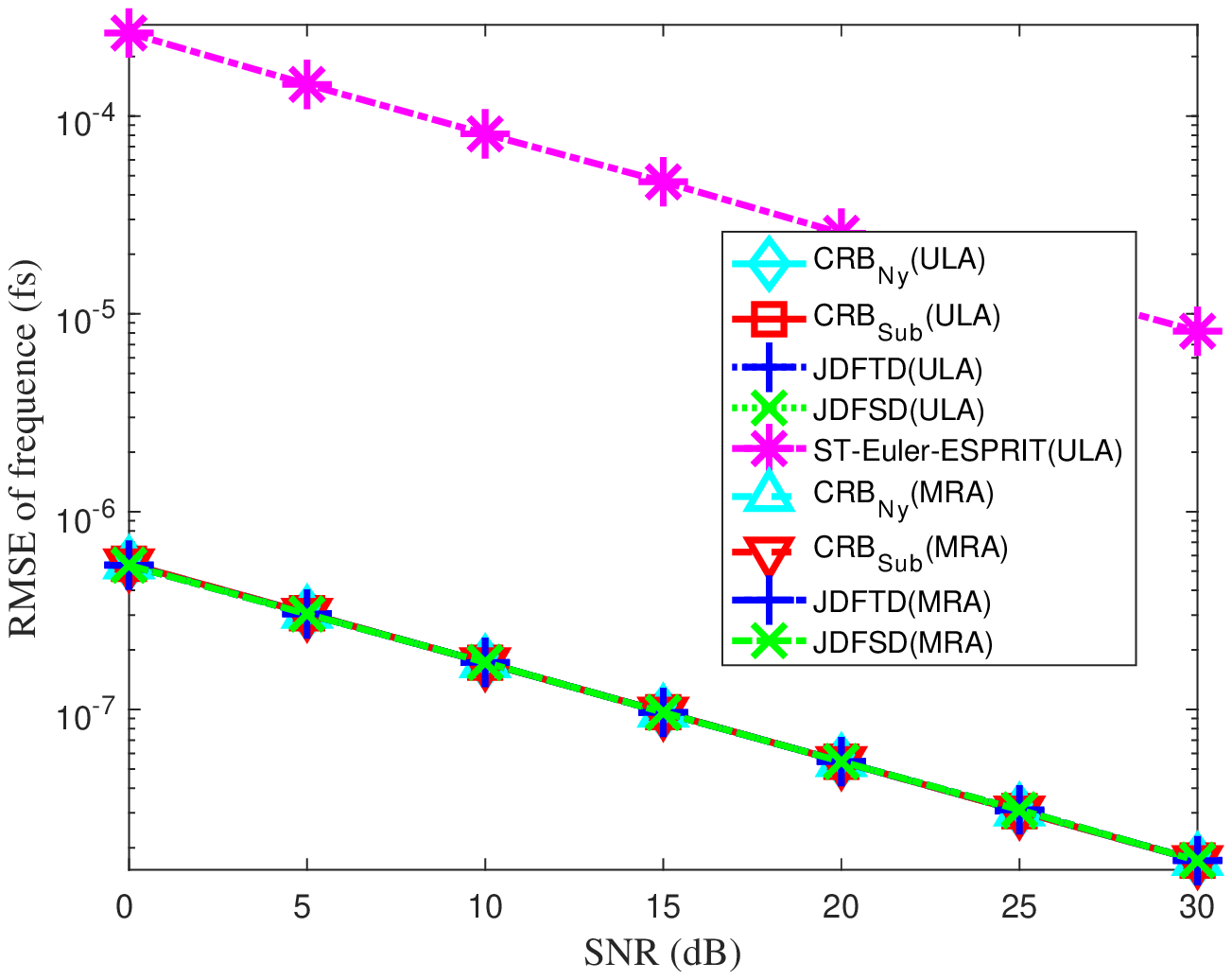}
\caption{RMSE of frequency estimates versus SNR.}
\label{figFS}
\end{figure}

\begin{figure}[!t]
\centering
\includegraphics[width=2.5in]{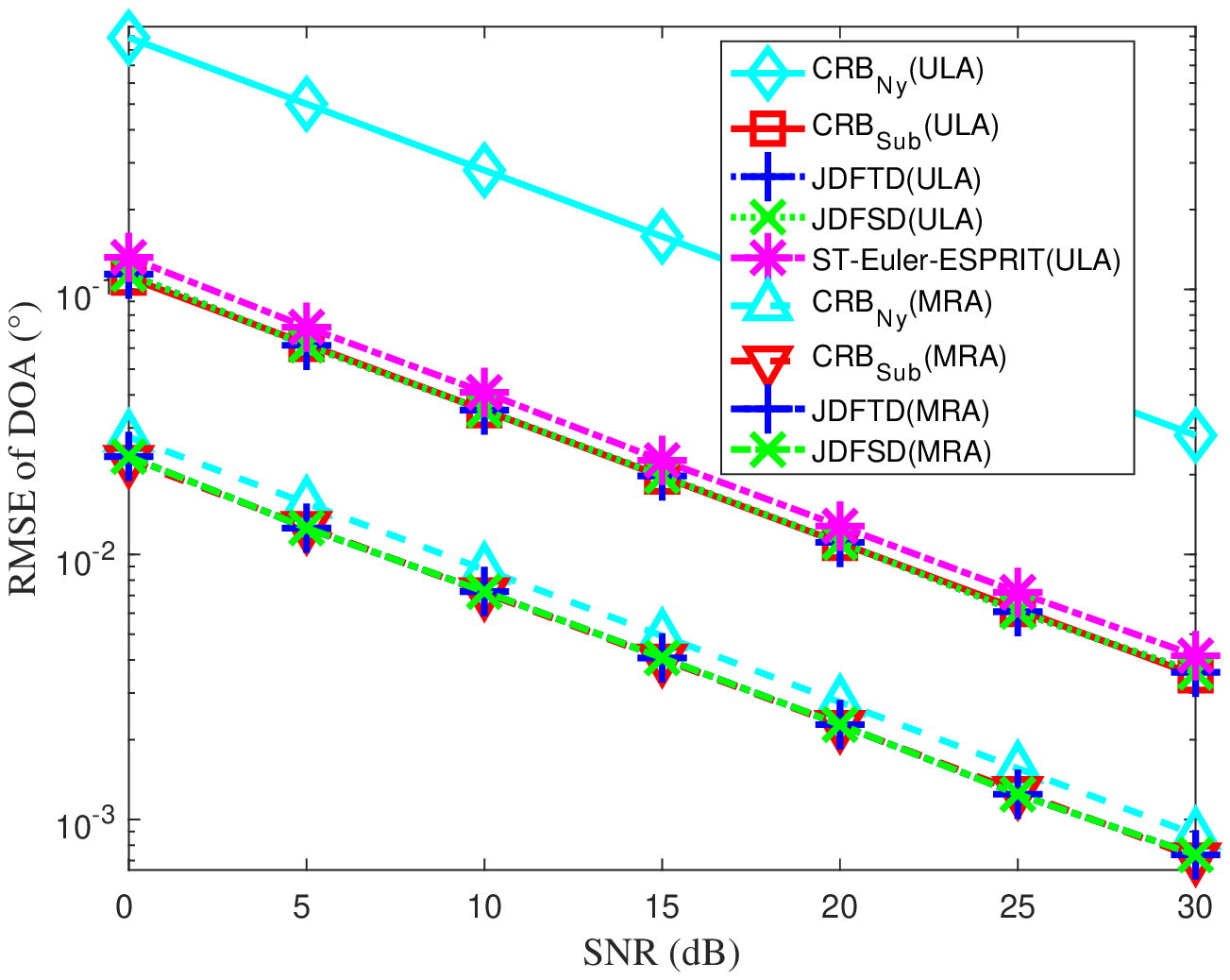}
\caption{RMSE of DOA estimates versus SNR.}
\label{figDS}
\end{figure}

Fig.\ref{figPS}-Fig.\ref{figDS} depict the RMSE versus  SNR in terms of spatial phase, frequency, and DOA  estimation, respectively.  Fig.\ref{figPS} shows that the phase estimation performance of algorithms JDFTD and JDFSD improves with SNR  and achieves the $\textrm{CRB}_{sub}$ when ULA or MRA is employed.
{Although ST-Euler-ESPRIT has a similar trend, the performance is inferior to JDFTD and JDFSD when ULA is employed. And we also find that the phase estimation performance of JDFTD and JDFSD  will improve obviously when MRA is employed since the MRA widens the array aperture. But ST-Euler-ESPRIT can not employ the MRA as explained previously.
As expected, Fig.\ref{figPS} shows that $\textrm{CRB}_{sub}$ is lower than $ \textrm{CRB}_{Ny}$ when $L=P$ and $K>1$. This illustrates that our algorithms for the new model can obtain a better phase estimation than the Nyquist sampling structure.}

Fig.\ref{figFS} demonstrates that the frequency estimation performance of JDFTD and JDFSD can achieve the $\textrm{CRB}_{sub}$. {But the frequency estimation performance of ST-Euler-ESPRIT is obviously worse than JDFTD and JDFSD.}
 From the view of frequency estimation, employing the array receiving is equal to enhancing the SNR. So the frequency estimation performance is the same when the sensor number is the same whether the array is  ULA or MRA. Combined with the conclusion in \cite{Rife1974},  the $\textrm{CRB}_{sub}$ and $\textrm{CRB}_{Ny}$ for frequency estimation are given as follow when $T$ is sufficiently large.

\begin{align}\label{CRBfre}
{\rm{CR}}{{\rm{B}}_{sub}}\left( f \right){\rm{ }} &= \frac{1}{{4\pi }^2}\frac{6}{{{\rm{SNR}}}}\frac{1}{M}\frac{{f_N^2}}{{{T^3}}}\frac{L}{P}.\\
{\rm{CR}}{{\rm{B}}_{Ny}}\left( f \right){\rm{ }} &= \frac{1}{{4\pi }^2}\frac{6}{{{\rm{SNR}}}}\frac{1}{M}\frac{{f_N^2}}{{{T^3}}}.
\end{align}
It is particularly obvious that  ${\rm{CR}}{{\rm{B}}_{sub}}\left( f \right)  = {\rm{CR}}{{\rm{B}}_{Ny}}\left( f \right)$ when $L=P$ according to (\ref{CRBfre}).

Comparing Fig.\ref{figDS} with Fig.\ref{figPS}, it is concluded that the performances of  DOA  estimation and phase estimation have the same trend. Because the sampling number $M$ in space domain is much less than the sampling number $T$ in time domain, the phase estimation is worse than the frequency estimation. Based on (\ref{Phi}), we know that the performance of DOA estimation is mainly influenced by the phase estimation. Because of this, we will only give the phase estimation simulation result rather than the DOA estimation simulation result in the following simulations.

\subsection{Performance with various branch number}
In this  subsection, we will investigate the estimation performance in the case of different branch number. The simulation conditions are the same with that in subsection \ref{SimuB} except that the branch number $P$ changes from 4 to 20 at 2 interval -  that is, the average sampling rate ${f_A} = \frac{{P{f_N}}}{L}$ changes from $0.2{f_N}$ to ${f_N}$ at $0.1{f_N}$ interval. In this simulation, we employ a random sampling pattern $C$.

\begin{figure}[!t]
\centering
\includegraphics[width=2.5in]{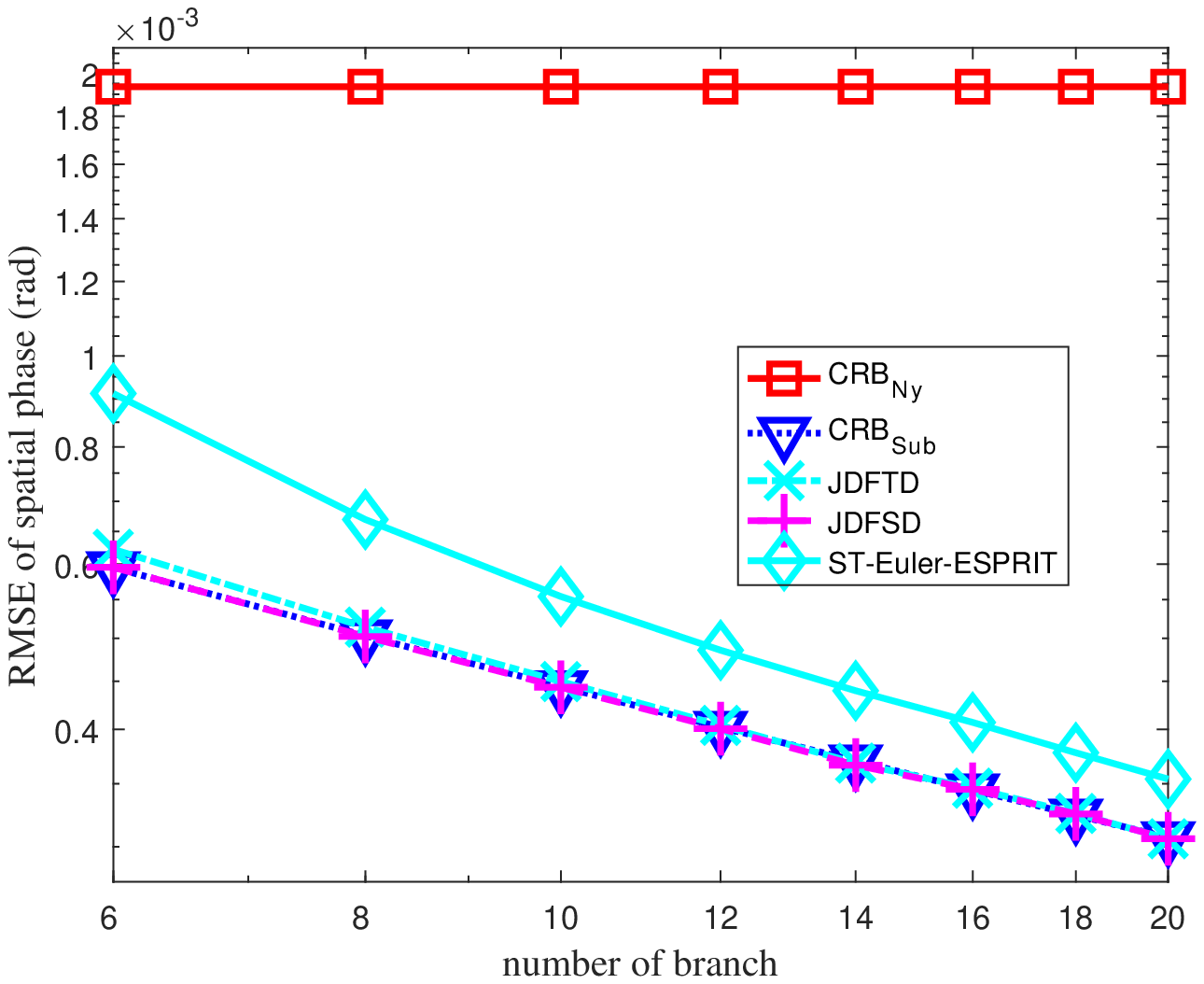}
\caption{RMSE of phase estimates versus number of branch.}
\label{figPC}
\end{figure}

Fig.\ref{figPC} shows that the phase (DOA) estimation performances of algorithm JDFTD and JDFSD improve with the branch number and reach the $\textrm{CRB}_{sub}$ when $P$ is relatively large. JDFTD is slightly worse than JDFSD or $\textrm{CRB}_{sub}$. This is because that the cross-correlation of the column vectors of $\bf{B}$ is obviously great when $P$ is {remarkably} small and this leads to trilinear decomposition slow convergence. It is not surprising that the $\textrm{CRB}_{sub}$ is lower than $\textrm{CRB}_{Ny}$ when $L=P$.  But we notice that this phenomenon still exists even $P=0.2L$. This illustrates that the benefit from the decrease of the cross-correlation of steer vectors from (\ref{CorB}) is much larger than the loss  caused by the decrease of samplings. Namely, we can realize a much better phase and DOA estimation performance with much less samplings. {Fig.\ref{figPC} also shows that both JDFTD and JDFSD are superior to ST-Euler-ESPRIT with any branch numbers, especially with small numbers.}

\begin{figure}[!t]
\centering
\includegraphics[width=2.5in]{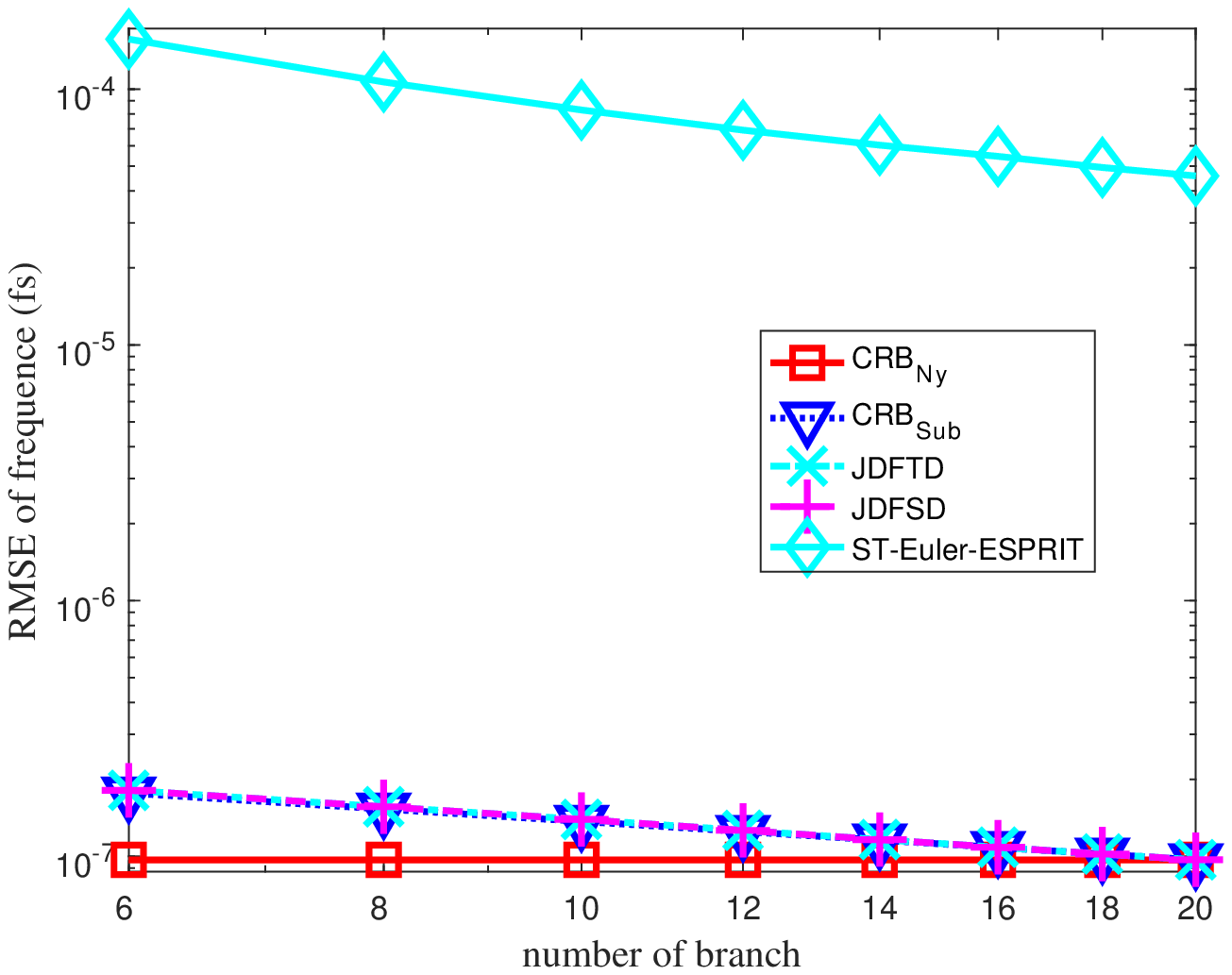}
\caption{RMSE of frequency estimates versus number of branch.}
\label{figFC}
\end{figure}

Fig.\ref{figFC} shows that the frequency estimation performances of algorithm JDFTD and JDFSD are improved with the branch number and achieve the $\textrm{CRB}_{sub}$ when $P$ is relatively large. The performances of JDFTD and JDFSD are slightly worse than the $\textrm{CRB}_{sub}$. Obviously,  $\textrm{CRB}_{sub}$ is higher than $\textrm{CRB}_{Ny}$ except $P=L$. {However, the frequency estimation performance of ST-Euler-ESPRIT is obviously worse than JDFTD and JDFSD no matter how many branches there are.}

We notice that when the average sampling rate is lower than the Nyquist sampling rate where $P<L$,  the CRB for our model is lower than the conventional CRB  in terms of DOA or spatial phase  estimation, and the opposite happens in terms of frequency estimation. However, the performance of DOA or spatial phase estimation is usually  far worse than that of frequency estimation because of $M \ll T$. So, we care more about the  performance improvement of DOA or spatial phase estimation than that of frequency estimation. Namely, comparing with the  performance improvement of DOA or spatial phase estimation, the   performance degradation of frequency estimation is insignificant.

\begin{figure}[th]
\centering
\includegraphics[width=2.5in]{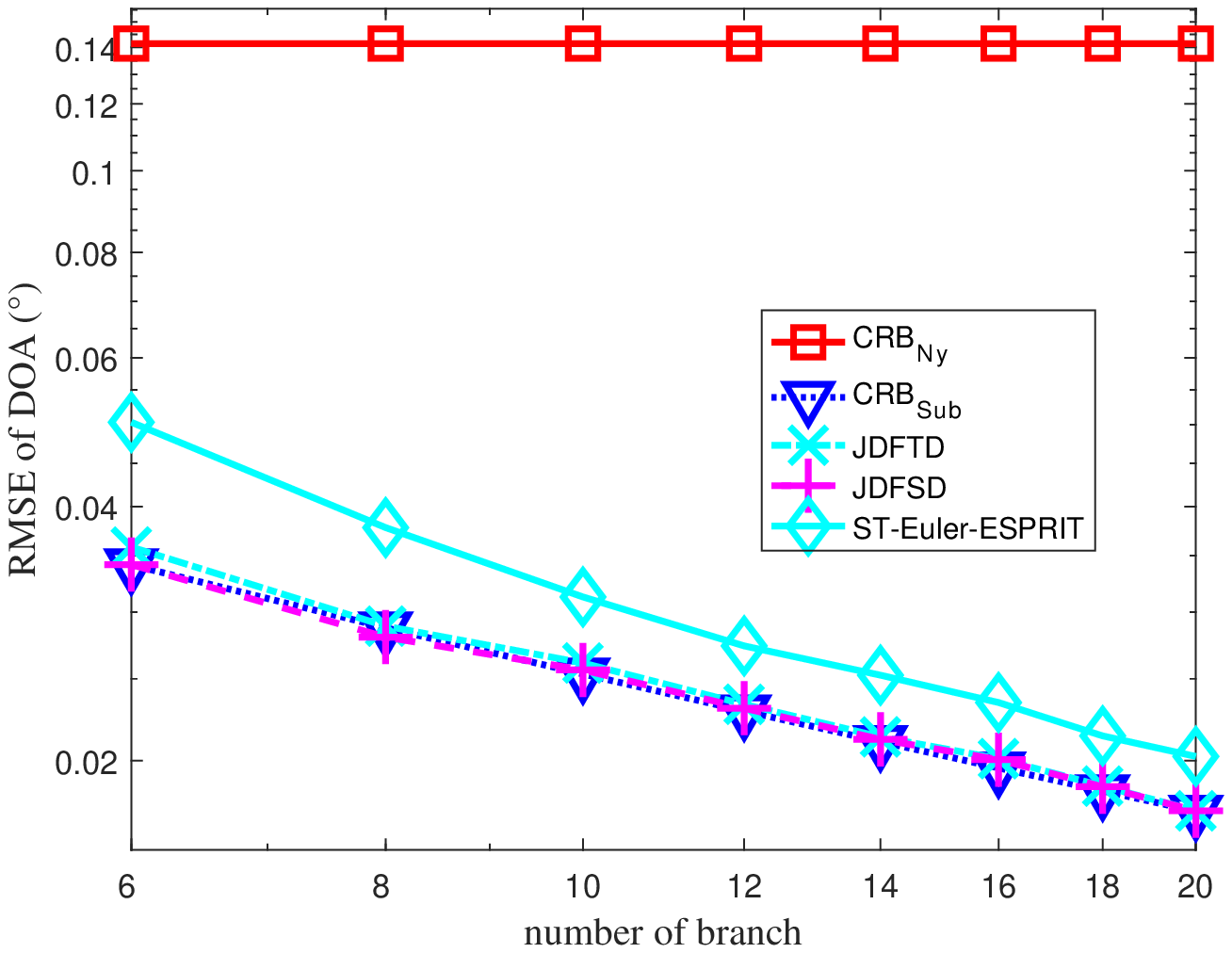}
\caption{RMSE of DOA estimates versus number of branch.}
\label{figDC}
\end{figure}

\subsection{Performance with various signal number}
In this  subsection, we will investigate the estimation performance when the signal number changes. The simulation conditions are the same with  subsection \ref{SimuB} except that the signal number $K$ changes from 1 to 5 one by one and only ULA is considered. We set ${\bm{\vartheta  }} = \left[ {{\vartheta _1},{\vartheta _2},\cdot\cdot\cdot,{\vartheta _5}} \right]$, where ${\vartheta _i}$ is subject to $[10j - {2.5^\circ },10j + {2.5^\circ }]$ uniformly distribution, where $i =  - 2, - 1, \cdots ,2$, $j =  - 2, - 1, \cdots ,2$ and $i$ has no corresponding relationship with $j$.  We set DOA ${\bm{\theta }} = \left[ {{\vartheta _1},{\vartheta _2},\cdot\cdot\cdot,{\vartheta _K}} \right]$. Let $\bm{\upsilon}  = \left[ {{\upsilon _1},{\upsilon _2}, \cdots ,{\upsilon _5}} \right]$ are subject to $[0.5,9.5]$ GHz uniformly distribution, at the same time any two signals are not in the same one sub-band. The signal carrier frequency is set at ${\bf{f}} = \left[ {{\upsilon_1},{\upsilon_2},\cdots,{\upsilon_K}} \right]$.

\begin{figure}[!t]
\centering
\includegraphics[width=2.5in]{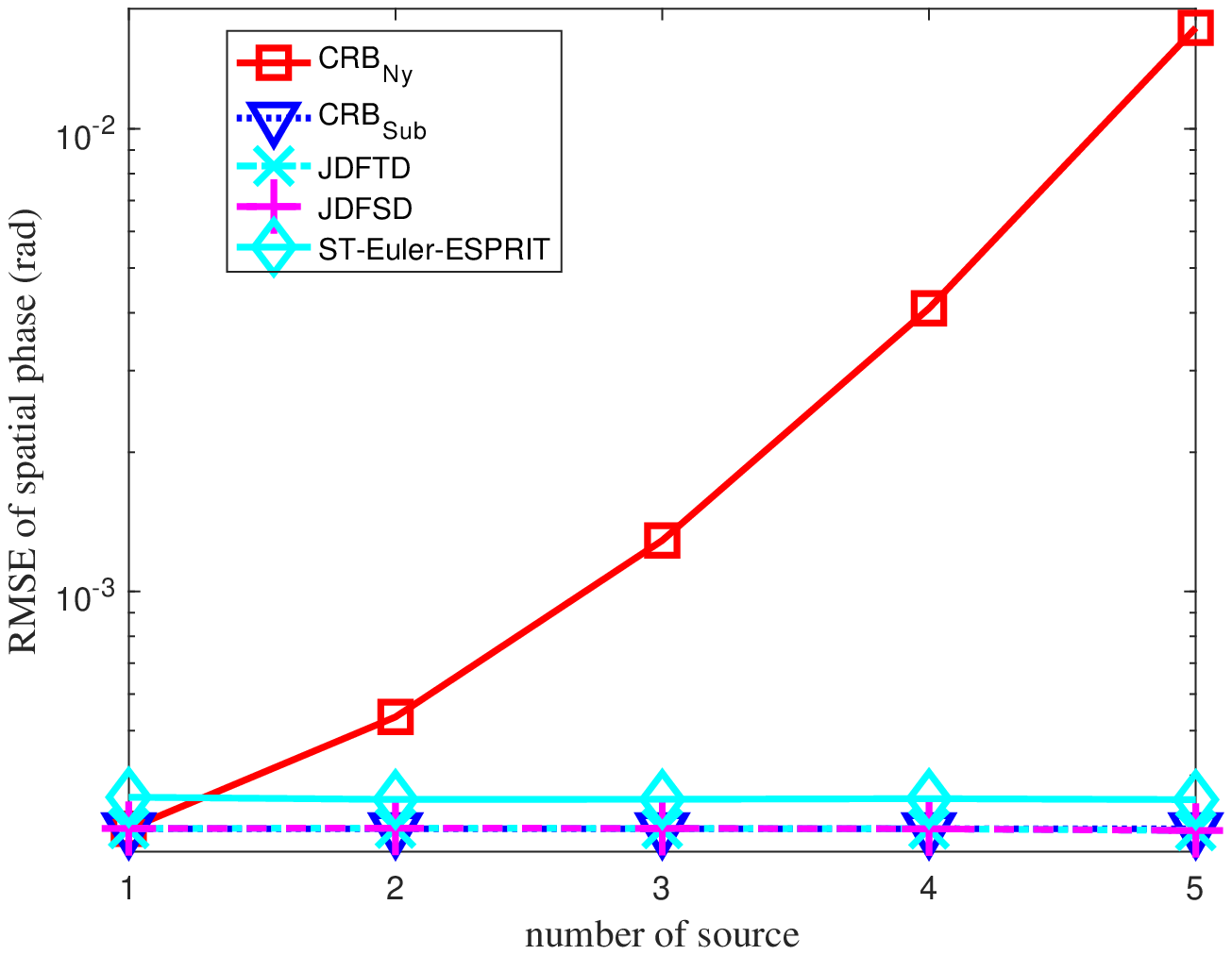}
\caption{RMSE of phase estimates versus number of source.}
\label{figPK}
\end{figure}

Fig.\ref{figPK} shows that the phase (DOA) estimation performances of algorithm JDFTD and JDFSD are not influenced by the signal number and maintain $\textrm{CRB}_{sub}$. However, $\textrm{CRB}_{Ny}$  increases with the signal number, and  increases faster than exponential function of the signal number. Besides, $\textrm{CRB}_{Ny}$ is equal to $\textrm{CRB}_{sub}$ only when $K=1$, {otherwise} $\textrm{CRB}_{Ny}$ is higher than $\textrm{CRB}_{sub}$. {These results} meet the analysis in section \ref{SectionCRB}. {As for ST-Euler-ESPRIT, although it is also negligible effected by the signal number, its phase (DOA) estimation performance is worse than JDFTD and JDFSD.}

\begin{figure}[!t]
\centering
\includegraphics[width=2.5in]{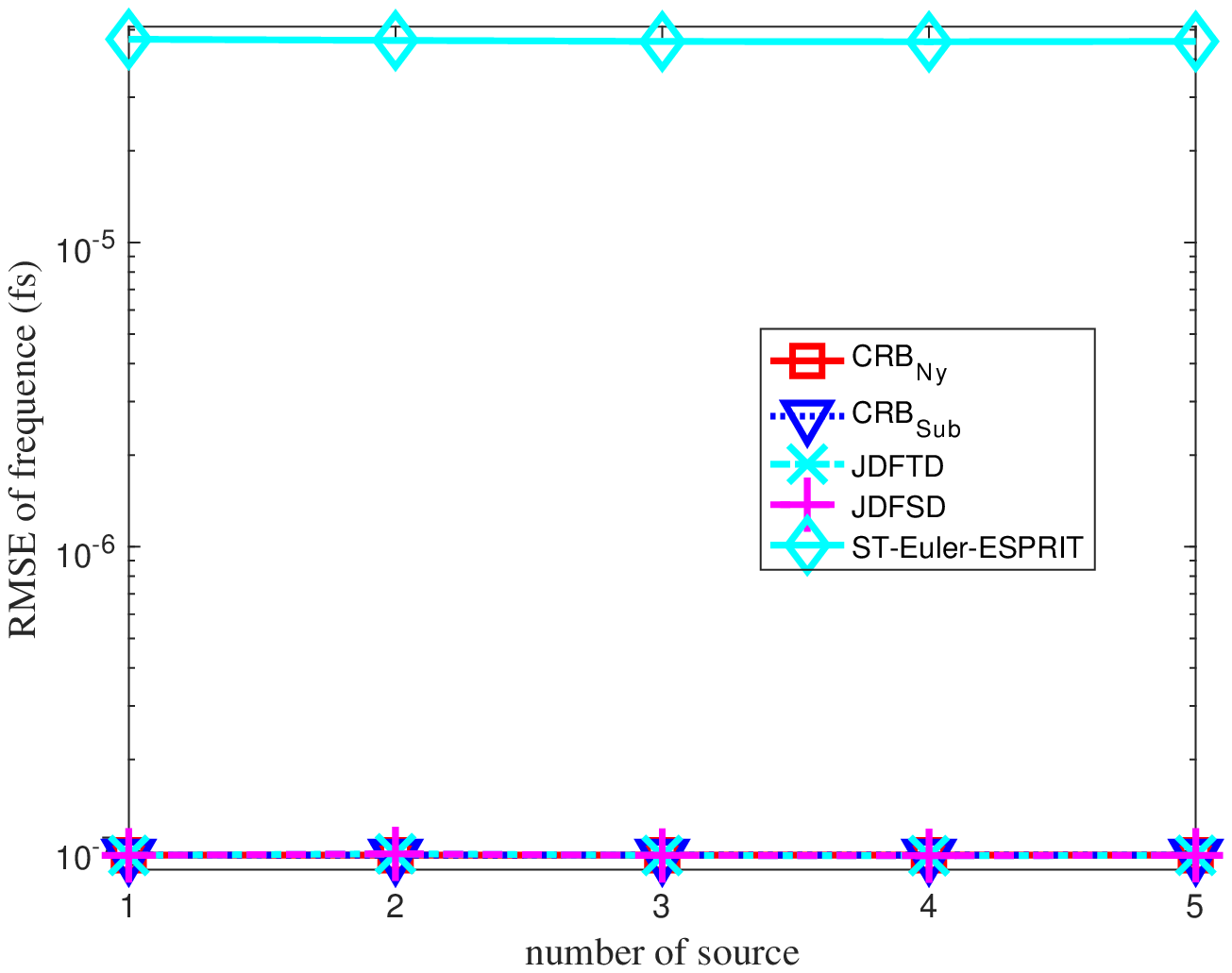}
\caption{RMSE of frequency estimates versus number of source.}
\label{figFK}
\end{figure}

Fig.\ref{figFK} shows that the frequency estimation performances of algorithm JDFTD and JDFSD are not influenced by the signal number and can reach  $\textrm{CRB}_{sub}$. At the same time, $\textrm{CRB}_{Ny}$ is equal to $\textrm{CRB}_{sub}$ because $P=L$. {The frequency estimation performance of ST-Euler-ESPRIT is particularly worse than that of JDFTD and JDFSD as before.}

\section{Conclusions}

In this paper, for the scenario where several narrowband far-field signals whose carrier frequencies are far separated impinging on an array, we designed an array receiver architecture by introducing the sub-Nyquist sampling technology. {We  derived a time-space union signal reception model with taking the spatial sampling and sub-Nyquist sampling into consideration simultaneously. Meanwhile, we can decrease the time-domain sampling rate and improve the DOA estimation accuracy.}

We proposed two joint DOA and frequency estimation algorithms for this model, one is based on trilinear decomposition from the perspectives of third-order tensor and the other is based on subspace decomposition.

In terms of spatial phase estimation, we derived the CRB for the model, and proved that the CRB is immune to the signal number when the branch number of our architecture is equal to the sampling rate reduction factor, and is lower than the CRB for the conventional model which employs Nyquist sampling. Furthermore, the new steer vectors are completely uncorrelated under the limited number of sensors, which makes a big improvement for the spatial phase estimation performance.  {From the geometry perspective, the estimation performance improvement benefits from the union time-space model.}

The simulations validated that the receiver architecture and the proposed approaches are feasible, and the variances of the proposed approaches are very close to their CRB and are beyond the CRB which employs Nyquist sampling in the case of different noise levels, different branch number, or different source number. Specifically, the variances of the proposed approaches are lower than the CRB which employs Nyquist sampling when the branch number of our architecture is less than the sampling rate reduction factor-that is, the average sampling rate is lower than the Nyquist sampling rate.

\ifCLASSOPTIONcaptionsoff
  \newpage
\fi

\bibliographystyle{IEEEtran}
\bibliography{IEEEabrv,Ref}

\end{document}